\documentclass{aa}

\def\ergscm2{\,erg\,cm^{-2}\,s^{-1}}
\def\ergs{\,erg\,s^{-1}}
\def\cm2{\,cm^{-2}}
\def\scm2{\,cm^{-2}\,s^{-1}}
\def\ss{\,s\,s^{-1}}

\def\xte{XTE J1810$-$197}

\def\ee {1E\,1048.1$-$5937}

\def\axj {AX\,J1844$-$0258}

\def\ea {1E\,2259$+$586}
\newcommand{\AXAF}{{\em Chandra}}
\input psfig.sty

\begin{document}

\title{Infrared  and X--ray variability of the transient Anomalous X-ray
Pulsar XTE J1810-197\thanks{Partially based on observations
carried out at ESO, Cerro Paranal, Chile (072.D--0297)}}

\author{N. Rea$^{1,2}$, V. Testa$^{2}$, G.L. Israel$^{2,3}$, S. Mereghetti$^{4}$, R.Perna$^{5}$, L. Stella$^{2,3}$,  A. Tiengo$^{4,6}$, \newline V. Mangano$^{7}$, T. Oosterbroek$^{8}$,  R. Mignani$^{9}$, G.Lo Curto$^{10}$, S. Campana$^{11}$, S. Covino$^{11}$}

\institute{ Universit\'a di Roma 2, Via della Ricerca Scientifica
1, I--00133 Roma, Italy: rea@mporzio.astro.it 
\and
INAF--Osservatorio Astronomico di Roma, Via Frascati 33, I--00040
Monteporzio Catone, Italy 
\and Affiliated with International Center for Relativistic Astrophysics 
\and CNR-IASF, Sezione di
Milano ``G.Occhialini'', Via Bassini 15, I--20133 Milano, Italy
\and Department of Astrophysical Sciences, Princeton University,
Princeton, NJ 08544 
\and Universit\`{a} degli Studi di Milano, 
Dipartimento di Fisica, v. Celoria 16, I-20133 Milano, Italy 
\and 
CNR-IASF, Sezione di Palermo, Via Ugo La Malfa 153, I--90146 Palermo, Italy
\and
Astrophysics Missions
Division, Research and Scientific Support Department of ESA,
ESTEC, Postbus 299, NL-2200 AG Noordwijk, Netherlands 
\and
European Southern Observatory, Karl--Schwarzschild str. 2,
D--85748 Garching, Germany \and European Southern Observatory, Av.
Alonso de Cordova 3107, Vitacura, Casilla 19001, Santiago 19,
Chile 
\and INAF --Osservatorio Astronomico di Brera, Via Bianchi
46, I--23807 Merate (Lc), Italy}
\date{}
\offprints{rea@mporzio.astro.it}
\authorrunning{Rea et al.}
\titlerunning{IR and X-ray variability of \xte}

\abstract {We report on observations aimed at searching for flux
variations from the proposed IR counterpart of the Anomalous X--ray
Pulsar (AXP) \xte. These data, obtained in March 2004 with the
adaptive optics camera NAOS-CONICA at the ESO VLT, show that the
candidate proposed by Israel et al. (2004) was fainter by $\Delta
H=0.7\pm0.2$ and $\Delta K_{s}=0.5\pm0.1$ with respect to October
2003, confirming it as the IR counterpart of \xte.  We also report on
an XMM--Newton observation carried out the day before the VLT
observations. The 0.5-10\,keV absorbed flux of the source was
$2.2\times10^{-11}\ergscm2$, which is less by a factor of about two
compared to the previous XMM--Newton observation on September 2003.
Therefore, we conclude that a similar flux decrease took place in the
X--ray and IR bands. We briefly discuss these results in the framework
of the proposed mechanism(s) responsible for the IR variable emission
of AXPs.}

\maketitle

\section{Introduction}

The X-ray source \xte\, was discovered in July 2003 as a transient
pulsar with a flux of $\sim5\times10^{-11}\ergscm2$ and a period
of 5.5 s (Ibrahim et al. 2004; Markwardt et al. 2003). Already
from the first RXTE and \AXAF\, results it clearly appeared that
the properties of \xte\, are different from those of the majority
of X-ray transient pulsars. The latter are easily identified as
neutron stars accreting from companion stars. On the other hand
the long term spin-down at $\sim10^{-11}\ss$, the soft X-ray
spectrum, and the upper limits on its optical counterparts
indicated \xte\, as a likely member of the class of Anomalous
X-ray Pulsars (AXPs, see Mereghetti et al. 2002 and 
Woods \& Thompson 2004 for a review). The
AXP nature of \xte\, was further strengthened by an XMM-Newton
observation (Tiengo \& Mereghetti 2003; Gotthelf et al. 2004)
showing the blackbody plus power law spectrum typical of this
class of sources, as well as by the identification of a candidate
IR counterpart with $K_{s}\sim20$ (Israel et al. 2004).

The AXP constitute an enigmatic class of pulsars, most likely hosting
young neutron stars, which has attracted increasing interest since its
first recognition (Mereghetti \& Stella 1995, van Paradijs et
al. 1995). The rotational energy loss inferred from their spin-down,
assuming they are neutron stars, is insufficient to power their X-ray
luminosity of $\sim10^{34}-10^{35}\ergs$, and they lack evidence of
companion stars which could power the emission through mass accretion.
Similarities with the persistent X-ray counterparts of the Soft
Gamma-ray Repeaters (e.g. Hurley 2000; Woods \& Thompson 2004) led to
the speculation that the AXP might be powered by the decay of strong
magnetic fields (Duncan \& Thompson 1992; Thompson \& Duncan
1995). Recent observations of bursts from the AXPs \ea\, and \ee\, and
support this ''Magnetar'' model (Kaspi et al. 2003; Gavriil, Kaspi \&
Woods 2002; Kaspi et al. 2004), but also other scenarios are discussed
by several authors and not completely ruled out. For example several
models based on accretion from residual disks around isolated neutron
stars have been proposed (Alpar 2001; Perna et al. 2000; Chatterjee et
al. 2000).

The particular interest for \xte\, lies in its transient nature,
as testified by archival observations showing a luminosity a
factor $\sim100$  below that observed in 2003-2004. In fact, most
AXP have shown until recently little or no long term
variability\footnote{the only exception being \axj\, (Gotthelf et
al. 2001) , which is actually a ''candidate'' AXP, detected only
once and for which a period derivative measurement is still
lacking}.
The results on \xte\, imply the existence of a, possibly large,
population of quiescent AXPs in the Galaxy, with relevant
implications on the nature and evolution of these objects. Recent
observations have also revealed variations in the X-ray flux of a
few ''persistent'' AXPs (Kaspi et al. 2003; Mereghetti et al.
2004; Gavriil \& Kaspi 2004), as well as in their infrared counterparts (see Israel et
al. 2004a). However, a clear picture of these variability
properties has still to emerge. In this context, multi-wavelength
monitoring of \xte\, , the only confirmed transient AXP, can yield
interesting results. Here we report on nearly simultaneous X-ray
and IR observations of \xte\,, showing a flux decrease in both
bands which confirms the proposed IR identification.

\section{Data analysis and results}

Deep IR imaging was obtained at the VLT-UT4 Yepun with the Nasmith
Adaptive Optics System and the High Resolution Near IR Camera
(NAOS-CONICA) on 2004 March 12, 13 and 14.  The pixel size of the
camera is 0.027$^{\prime\prime}$.
Images were reduced with the instrument-specific pipelines and
checked by reducing them again with the software package {\em
eclipse}. A total of 18 cube images in Ks and 26 in H, of 40\,s
exposure each, were obtained, for a total exposure time of 36 and
52 minutes in Ks and H, respectively. The on-axis FWHM was
determined to be $0.09^{\prime\prime}$ (3.3 pixels) in $K_{s}$ and
$0.10^{\prime\prime}$ (3.6 pixels) in H.

Aperture photometry was performed with the {\em digiphot} package
of IRAF\footnote{task {\em phot} in the {\em daophot} package
(Stetson 1990), and corrected for aperture at infinity with the
{\em mkapfile} routine available in the package {\em photcal}}.
The two output catalogs were matched and calibrated by using a set
of secondary standards in the field, for which magnitudes and
colors were already obtained in October 2003 (Israel et al. 2004).
We found the following IR magnitudes for the proposed IR
counterpart: $K_{s}=21.36\pm0.07$ and $H=22.73\pm0.18$. The object
was not detected in the J band (J$>$23.0, 3$\sigma$ u.l.). The
corresponding values in October 2003 were $K_{s}=20.8\pm0.1$ and
$H=22.0\pm0.1$.

As shown in Fig.\,1, where the difference between the  magnitudes
in October 2003 and March 2004 are plotted for all the objects
within $\sim$7\arcsec\ from the  Chandra AXP position, only the
proposed IR counterpart to \xte\ showed a significant variation,
with $\Delta H=0.7\pm0.2$ and $\Delta K_{s}=0.5\pm0.1$.


\begin{figure}[ht] 
\centerline{\psfig{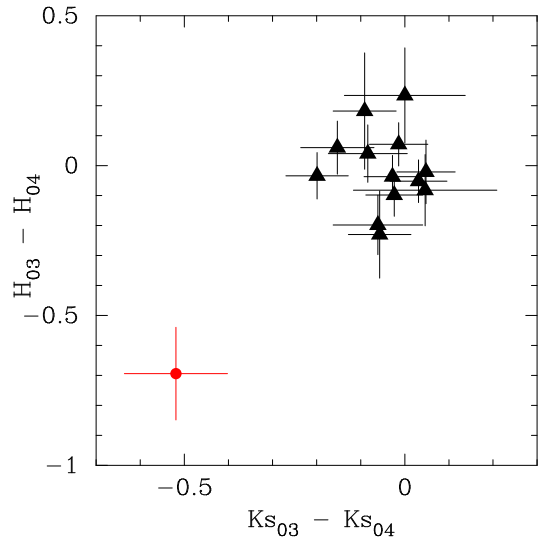} } 
\caption{VLT NAOS--CONICA H and $K_{s}$ band magnitude differences
between October 2003 and March 2004 observations obtained for all the
objects detected within a circular region of about 7'' radius from the
Chandra position (Israel et al. 2004). Filled circle represents the
\xte\ measurements, while filled triangles the differences for all the
other stars.}
\end{figure} 


\xte\, was observed with XMM--Newton for  16 ks on 11 March 2004.
All the EPIC cameras were used with the medium thickness filter
(Turner et al. 2001, Str\"{u}der et al. 2001). The MOS detector
was in Small Window mode (time resolution 300 ms over a 
$\sim$1.5$'\times$1.5$'$ field)
while the PN was in Large Window Mode (time resolution 48 ms over a
$\sim$13$'\times$26$'$ field). Standard SAS 6.0 tools were used for the data
reduction.
Coherent pulsations at 5.53\,s were detected with standard Fast
Fourier algorithms and the best period value of
$P=5.539917\pm0.000005$\,s (90\% confidence level; at the epoch:
53075.49196187 MJD) was obtained with phase fitting
techniques. Compared with the period measured in the September 2003
XMM--Newton observation, this yields an average $\dot{P} =
(3.6\pm0.1)\times10^{-11}$s s$^{-1}$. Comparing the frequency
derivative reported for the RXTE data of \xte\,(Ibrahim et al. 2004)
we found that a variation occurred in the period derivative, which is
now greater by a factor of about three respect to the value reported
for the July-September 2003 time span. The pulsed fraction
(semiamplitude of modulation divided by the mean source count rate) in
the 0.6--10\,keV energy range was $49\pm1$\%.

The source spectrum was extracted from a 32$''$ radius circle and the
background from source free regions in the field.  The spectrum was
well fitted by a two component model composed of an absorbed blackbody
plus a power law with
$N_{H}=0.96\pm0.03$\,$\times10^{22}$\,atoms\,$\cm2$,
kT=$0.67\pm0.01$\,keV and photon index of $\Gamma=3.8\pm 0.1$
($\chi^{2}_{\nu}=1.15$; uncertainties are at 90\% confidence level;
see Fig.2). The corresponding blackbody radius was $1.23\pm0.02$\,km,
for an assumed distance of $4$kpc (the estimated distance is 3--5 kpc;
Gotthelf et al. 2004). The 0.5-10\,keV absorbed flux was
$(2.2\pm0.1)\times10^{-11}\ergscm2$, corresponding to an unabsorbed
flux of $8.2\times10^{-11}\ergscm2$. The blackbody component accounts
for the 60\% of the absorbed flux in the 0.5-10\,keV band.

An equally acceptable fit ($\chi^{2}_{\nu}=1.23$) was also
obtained keeping the absorption fixed at the September 2003 value
($1.05\times10^{22}$\,atoms\,$\cm2$; Tiengo \& Mereghetti 2003, Gotthelf
et al. 2004). The resulting spectral parameters were
kT=$0.68\pm0.01$\,keV and $\Gamma=4.1\pm0.1$.

Using two blackbodies to fit the spectra we found: $N_{H}=
0.58\pm0.02$\,$\times10^{22}$\,atoms\,$\cm2$, $kT_{1}
=0.29\pm0.01$\,keV (radius of $5.1\pm0.6$\,km) and $kT_{2}
=0.70\pm0.01$\,keV (radius of $1.21\pm0.04$\,km;
$\chi^{2}_{\nu}=1.17$). The first blackbody had a 0.5-10\,keV
absorbed flux of $3.9\times10^{-12}\ergscm2$ (18\% of the total
flux) and the second of $1.8\times10^{-11}\ergscm2$.
Further analysis of the XMM-Newton observations, including phase
resolved spectroscopy, will be reported elsewhere.


\begin{figure}[htb] 
\centerline{\psfig{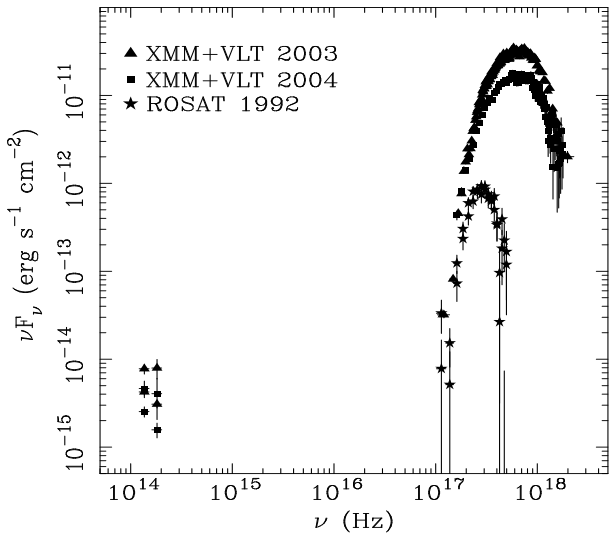} } 
\caption{Broadband energy spectrum of \xte. Filled squares represent the 2004 
XMM-Newton PN and VLT-NACO observations while triangles are relative
to the 2003 observations. Moreover filled stars represent the spectrum
of the ROSAT 1992 observation. Reported IR fluxes are absorbed and
unabsorbed.}
\end{figure} 


\section{Discussion}

Our new diffraction limited VLT images of the \xte\ field, carried
out six months after the previous ones, clearly show a decrease in
the IR flux of the candidate previously proposed based only on
positional coincidence and unusual colors (Israel et al. 2004).
This finding confirms it as the IR counterpart of \xte\ , which
showed during the same period a  similar variation in its X--ray
flux.

The only other case of correlated X-ray and IR flux variations in
an AXP observed to date was found after the detection of a series
of short bursts from \ea\ (Kaspi et al. 2003). In this case the
X-ray and IR fluxes decreased by a factor less than two in about
one week. The X--ray pulse shape, the pulsed fraction and the
spectral parameters changed significantly (Woods et al. 2003), and
a glitch was also observed. The  X-ray and IR  variability
reported here for \xte\, is not obviously tied to bursting activity
from the source (although the occurrence of bursts before the
observation, or at the time of the start of the outburst between
November 2002 and January 2003, cannot be excluded).


\begin{figure}[htb] 
\centerline{\psfig{figure=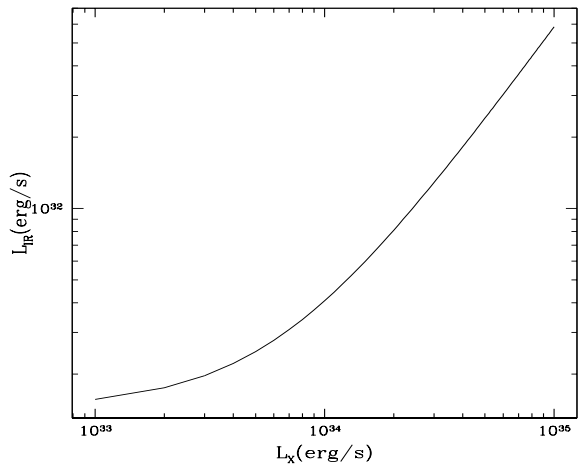,width=8cm,height=6.5cm} } 
\caption{Resulting correlation between X-ray and IR luminosity from a disk (assuming
a typical 60 deg inclination of the disk and a distance of $\sim$5\,kpc). }
\end{figure} 


Contrary to the case of \ea, we do not see large variations in the
timing or spectral properties of \xte\ between the two XMM-Newton
observations. There is only some evidence for a moderate softening of
the X--ray spectrum, as indicated by the change of the photon index
from $3.7\pm0.2$ (Gotthelf et al. 2004) to $4.1\pm0.2$ (for the
blackbody plus power-law model fits with a constant absorption).

Comparing the X-ray outburst of \xte\, with that recently found in
another AXP, \ee\, (Gavriil \& Kaspi 2004), we noticed that in both
cases the corresponding fluences are of the order of few
$10^{42}\ergs$. Moreover, looking at the published decaying
lightcurves for \xte\, and \ee\, (Ibrahim et al. 2004; Gavriil \&
Kaspi 2004), it is evident the similarity of the decaying law
behaviour, suggesting that both outbursts obay to the same physical
process.

The IR emission from AXP in the context of the magnetar model has been
recently discussed by \"{O}zel (2004), who noticed that the IR
emission cannot be due to thermal emission from the neutron star
surface and is consistent with synchrotron emission in the
magnetosphere. The value of the IR to X--ray flux ratio we derived for
\xte\ , when plotted versus the neutron star spin down luminosity,
does not follow the trend of the other AXPs shown in Fig.3 of \"{O}zel
(2004). However, for what concerning the other possibility proposed by
\"{O}zel (2004) in which the magnetospheric emission is powered by the
magnetic energy, no quantitative predictions are reported therefore we cannot
exclude this possibility.

The somehow correlated IR/X-ray fluxes of XTE J1810-197 can be accounted for
in a ``hybrid'' model of a magnetar surrounded by a fossil disk (Eksi
\& Alpar 2003).  The spectral characteristics of fall-back disks
around isolated neutron stars were studied in detail by Perna et
al. (2000) and Perna \& Hernquist (2000). They considered the
contribution to the emission from both viscous dissipation and
reprocessing of the X-ray luminosity from the star, finding that the
long wavelength emission, and in particular the IR, is dominated by
reprocessing of the X-ray radiation (presumably coming from a
magnetar).  This immediately implies that X-ray and IR flux variations
must be correlated.

We studied the extent of this correlation for an X-ray luminosity
on the order of a few $\times 10^{34}\ergs$ (assuming here a distance of 5
kpc) and for a disk model as described in the references above. We
found (see Fig.\,3) that a variation in X-ray luminosity of the
star by a factor of $\sim 2$ results in a corresponding variation
of the IR flux from the disk by also a factor of 2, and the
intensity of the predicted IR flux is also consistent with the
observations (assuming a typical 60$^{\circ}$ inclination of the
disk). These results are largely independent of the inner and
outer radius of the disk, as the IR emission is produced in a
small ring which, for the range of X-ray luminosities under
consideration is at a distance of a few $\times 10^{10}- 10^{11}$
cm.

In this scenario we expect $L_{IR}$ and $L_X$ to be correlated, even
if not linearly. This is due to the fact that, as $L_X$ increases, the
overall temperature in the disk consequently increases, and the region
with temperatures at which the IR radiation is produced moves towards
larger radii. This results in a larger emission area.  On the other
hand, as $L_X$ decreases, the flux from the disk becomes gradually
more dominated by viscous dissipation up to a point where this
completely takes over and $L_{IR}$ becomes independent of $L_X$:
however the $L_X$ limit value for the reprocessing dominated IR
emission is largely lower than the typical AXPs X-ray luminosity, we
then conclude that in the AXPs case the IR emission is almost
completely due to the X-ray reprocessing phenomenon. By considering
the AXPs sample as a whole, we would expect that X-ray brighter
objects would generally have brighter IR counterparts. Indeed, this
has been hinted at by Hulleman et al. (2004).

Our suggestion can be tested by a search for pulsations in the IR
radiation similarly to the search that Kern \& Martin (2002) performed
in the optical.

\end{document}